\documentclass[preprint,aps,prl,amssymb,rotating,epsfig,graphics,eepic,showpacs]{revtex4}
\usepackage{epsfig,psfig,rotating,color}
\usepackage{eepic,graphics,graphicx}
\newcommand{\be}{\begin{equation}} \newcommand{\ee}{\end{equation}}
\newcommand{\bea}{\begin{eqnarray}} \newcommand{\eea}{\end{eqnarray}}

\begin{document}

\title{Transverse fluctuations of grafted polymers}

\author{G. Lattanzi$^1$
  T. Munk$^{2}$ and E. Frey$^{1,2}$}

\affiliation{$^1$Abteilung Theorie, Hahn-Meitner-Institut, Glienicker
  Strasse 100, 14109 Berlin, Germany \\
  $^2$Fachbereich Physik, Freie Universit\"at Berlin, Arnimallee 14,
  14195 Berlin, Germany}

\date{\today}

\bibliographystyle{apsrev}

\begin{abstract}
We study the statistical mechanics of grafted polymers of  
arbitrary stiffness in a two-dimensional embedding space with 
Monte Carlo simulations. The probability distribution function of 
the free end is found to be highly anisotropic and non-Gaussian 
for typical semiflexible polymers. The reduced distribution in the 
transverse direction, a Gaussian in the stiff and flexible limits, 
shows a double peak structure at intermediate stiffnesses. We also 
explore the response to a transverse force applied at the polymer 
free end. We identify F-Actin as an ideal benchmark for the 
effects discussed. 
\end{abstract}


\maketitle
 
Healthy cells require an efficient and complex transport network 
to carry out the overwhelming number of tasks that are needed to 
accomplish their function. This network, also known as the {\em 
cytoskeleton}, is formed primarily by {\em filaments} (actin 
filaments, microtubules and intermediate filaments), linked 
together by a large collection of accessory 
proteins~\cite{howard}. A complete description of the structural 
and mechanical properties of these filaments is therefore 
essential in order to unveil the mechanical properties of the 
entire cell. Advances in the field have been significantly 
promoted by a unique set of optical and mechanical techniques 
which allow to visualize and manipulate single cytoskeletal 
filaments~\cite{ott93,git93,kae94} and DNA~\cite{mar95}. 
Fluorescence videomicroscopy~\cite{osk02} and 
nano-manipulation~\cite{liu02} can be conveniently used to obtain 
quantities as the distribution function of the end-to-end 
distance~\cite{osk02} or the mechanical response to an external 
force in great detail and at the single molecule level. These 
quantities are amenable to a direct comparison with theoretical 
models. 

The main material parameter in the description of a polymer 
filament is its persistence length, $\ell_p$. It is defined as the 
typical length over which correlations of the tangent vectors of 
the filament contour decay. Polymers are considered to be flexible 
when their persistence length is small compared to their total 
length $L$, or $t :=L/\ell_p \gtrsim 10$. In this limit, they can 
be well described by the minimal model of the Gaussian 
Chain~\cite{yamakawa}.  Polymers of biological importance, e.g. 
F-actin, are often semiflexible, meaning that their persistence 
length is comparable to their total length. While flexible 
polymers are dominated by entropic effects, the statistical 
mechanics of semiflexible polymers is strongly affected by their 
bending energy and the close vicinity of the classical Euler 
instability for buckling a rigid beam~\cite{lan7}. 

The distribution function $P(\vec R)$ of the end-to-end vector 
$\vec R$, a simple Gaussian for a flexible polymer, is peaked 
towards full stretching and is completely 
non-Gaussian~\cite{jan96}. The mechanical response of a 
semiflexible polymer is highly anisotropic, depending on the 
direction in which the force is applied~\cite{kro96}. These 
findings result in bulk properties of solutions and networks that 
are completely different from the isotropic elasticity of flexible 
polymer solutions~\cite{mac03,jan03}. In addition, the 
inextensibility constraint becomes crucial in determining the 
approach to full stretching upon the application of a force $f$, 
as reported by Marko and Siggia~\cite{mar95} for double--stranded 
DNA. 


Here we investigate the mechanical and statistical properties of a 
single chain grafted at one end, a problem of direct relevance for 
force generation in cellular systems. The other end is either 
free, or subject to a constant transverse force, whose magnitude 
extends into the non--linear regime.  We restrict ourselves to a 
two--dimensional embedding space, since in most experiments, 
fluctuations in one direction are severely restricted, or cannot 
be observed. The generalization to a three--dimensional space is 
straightforward and will be reported elsewhere~\cite{gia:unpub}. 

We refer to the Wormlike Chain Model (WLC) introduced by Kratky and
Porod~\cite{kra49}. In this framework, a polymer conformation is
represented by a succession of $N$ segments $\vec{t}_i$, whose
direction is tangent to the polymer contour at the $i$th segment.
Since the polymer is assumed to be inextensible, all segments
$\vec{t}_i$ have a prescribed length $a=L/N$. The Hamiltonian is given
by:
\be {\mathcal{H}} = -\varepsilon \sum_{i=1}^{N-1} \vec{t}_i \cdot
\vec{t}_{i+1} - \sum_{i=1}^N \vec{f} \cdot \vec{t}_i \, ,
\label{eq:discrete} 
\ee 
where $\varepsilon $ is the energy associated to each bond and
$\vec{f}$ is a force eventually applied to the second end. It is 
also possible to define a continuum limit for $a \to 0$, $N \to 
\infty$, with $N a = L$ and $\epsilon=\varepsilon a^2/N$ held 
fixed. The Hamiltonian in Eq.~\ref{eq:discrete} is then equivalent 
to the following functional~\cite{sai67,win94}: 
\be {\mathcal{H}}_f = \frac{\kappa}{2} \int_0^L ds \left(
  \frac{\partial \vec{t}(s)}{\partial s} \right)^2 - \vec{f} \cdot
\int_0^L ds \ \vec{t}(s) \, , \label{eq:hf} \ee
where $\kappa = \epsilon L$ and $\vec{t}(s)$ is the tangent vector 
of the space curve $\vec{r}(s)$ parametrized in terms of the arc 
length $s$. The inextensibility of the filament is imposed by the 
local constraint $\left| \vec{t} (s) \right| = 1$. The continuous 
version of the wormlike chain has been successfully used to obtain 
various statistical quantities, as the tangent-tangent correlation 
function or moments of the end-to-end distance 
distribution~\cite{sai67,nor78}. It has been recently used to 
obtain the radial distribution function~\cite{jan96}, and 
force-extension relations~\cite{mac95,mar95,kro96}. 

We have developed a Monte Carlo simulation to investigate the 
behavior of a semiflexible polymer in the proximity of the limit 
$t\to 1$. The rationale behind this choice is the search for clear 
hallmarks of the onset of the ``semiflexible'' nature of a 
filament. In this intermediate limit, analytical results are 
difficult to obtain: typical approximation schemes that build on 
either Gaussian chains or rigid rods are outside their validity 
range; hence, computer simulations become crucial. The first end 
of the filament is assumed to be clamped, i.e. the orientation of 
its tangent vector is held fixed along a direction, named the 
$x$-axis. The second end is left free to assume any possible 
orientation. The initial configuration has been randomly chosen in 
the proximity of the full stretching condition, thus ensuring a 
fast convergence to equilibrium. A new configuration is generated 
by changing the orientation of one segment and accepted according 
to the standard Metropolis algorithm and the discrete Hamiltonian, 
Eq.~\ref{eq:discrete}. Effects resulting from self--avoidance are 
not considered, but we notice that configurations where the chain 
folds back onto itself are strongly energetically suppressed for 
sufficiently stiff polymers. Results ceased to depend on the 
number of segments for $N=50$. On the order of $10^6$ Monte Carlo 
steps per segment were performed, and results were averaged over 
different runs, obtaining a perfect agreement between measured 
expectation values of the end-to-end distance $\left< R^2 \right>$ 
and $\left< R^4 \right>$ with known exact expressions. 
The radial distribution function was calculated and coincided with 
the analytic results in~\cite{jan96} within the accuracy thereby 
reported.

Here we are interested in the probability distribution function
$P(x,y)$ of the free end in the plane determined by the direction 
of the clamped end ($x$-axis) and the transverse one ($y$-axis). 
This quantity is directly accessible to experiments allowing for a 
quantitative comparison with our predictions. We will also 
consider the reduced distribution functions $P(x)$ and $P(y)$, 
obtained by integrating $P(x,y)$ over the variables $y$ and $x$, 
respectively. 

It is important to notice that when both ends are free, the radial
distribution function is rotationally invariant and is therefore 
only a function of the distance $R$ between the ends. Clamping one 
end breaks rotational symmetry and leads to distinctly different 
longitudinal and transverse distribution functions, $P(x)$ and 
$P(y)$. Nonetheless, the broken rotational symmetry does not 
affect the total energy of the configuration. This implies, and is 
in fact confirmed by our simulations (data not shown), that the 
longitudinal distribution function $P(x)$ coincides with the 
radial distribution function $P(R)$ of the end-to-end distance, 
apart from a constant normalization factor. The characteristic 
feature of this function is a crossover from a universal Gaussian 
shape centered at the origin with a characteristic width 
determined by the radius of gyration, to yet another universal 
shape~\cite{jan96}, whose peak is shifted towards full stretching 
and whose width is determined by a new longitudinal length scale 
$L_\parallel \propto L^2/\ell_p$. 

This has to be contrasted with the transverse distribution 
function. Not surprisingly, given the intrinsic isotropy of 
flexible polymers, the distribution $P(y)$ is a Gaussian and 
identical to $P(x)$ for high values of $t$. In the stiff limit, 
$P(y)$, at variance with $P(x)$, is again a Gaussian centered at 
$y=0$, whose width is now given by a new transverse length scale 
$L_\perp = \sqrt{2 L^3/3 
  \ell_p}$~\cite{wilhelm:unpub,ben03}. Surprisingly, at intermediate values
the probability distribution function is not a smooth interpolation
between these two Gaussian limits but shows interesting and
qualitatively new features.
\begin{figure}[htb!]
  \epsfig{figure=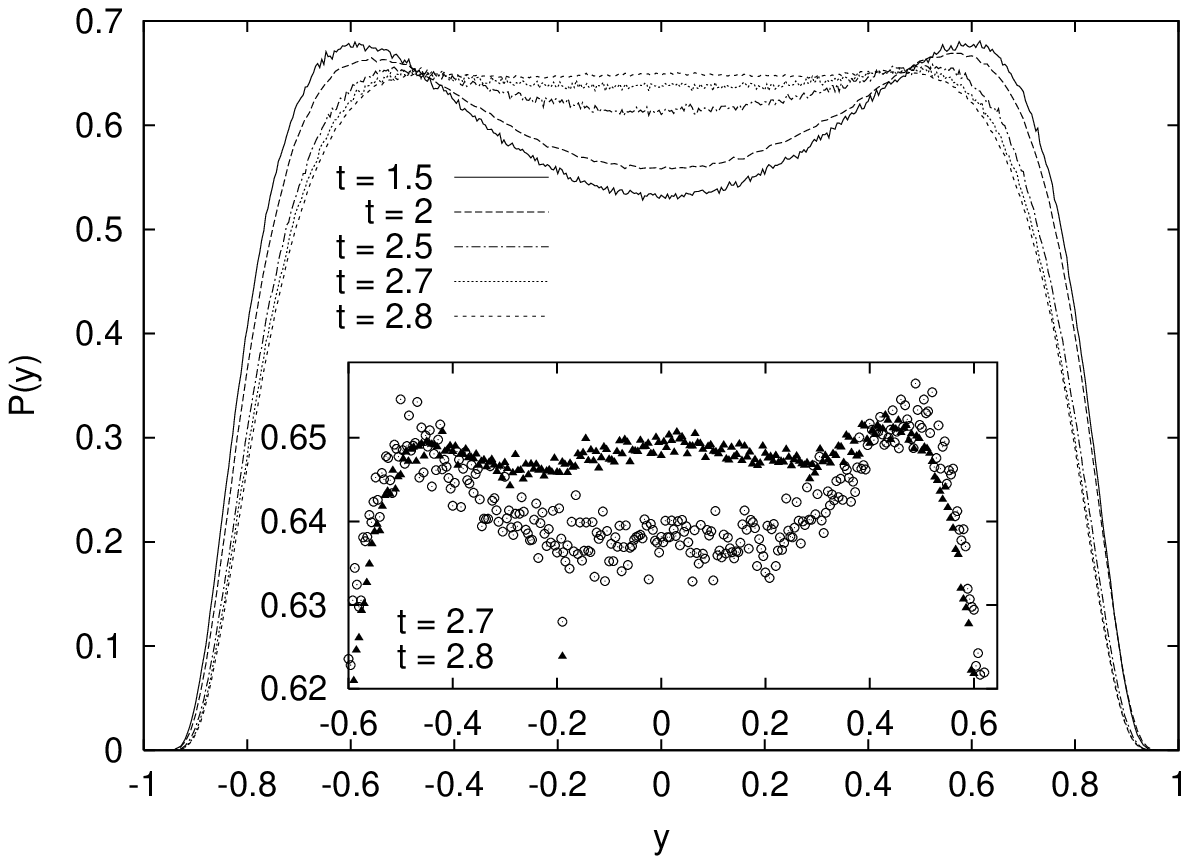,width=8.5cm,clip=} \put(-210,150){(a)}
  \put(-242,-165){\epsfig{figure=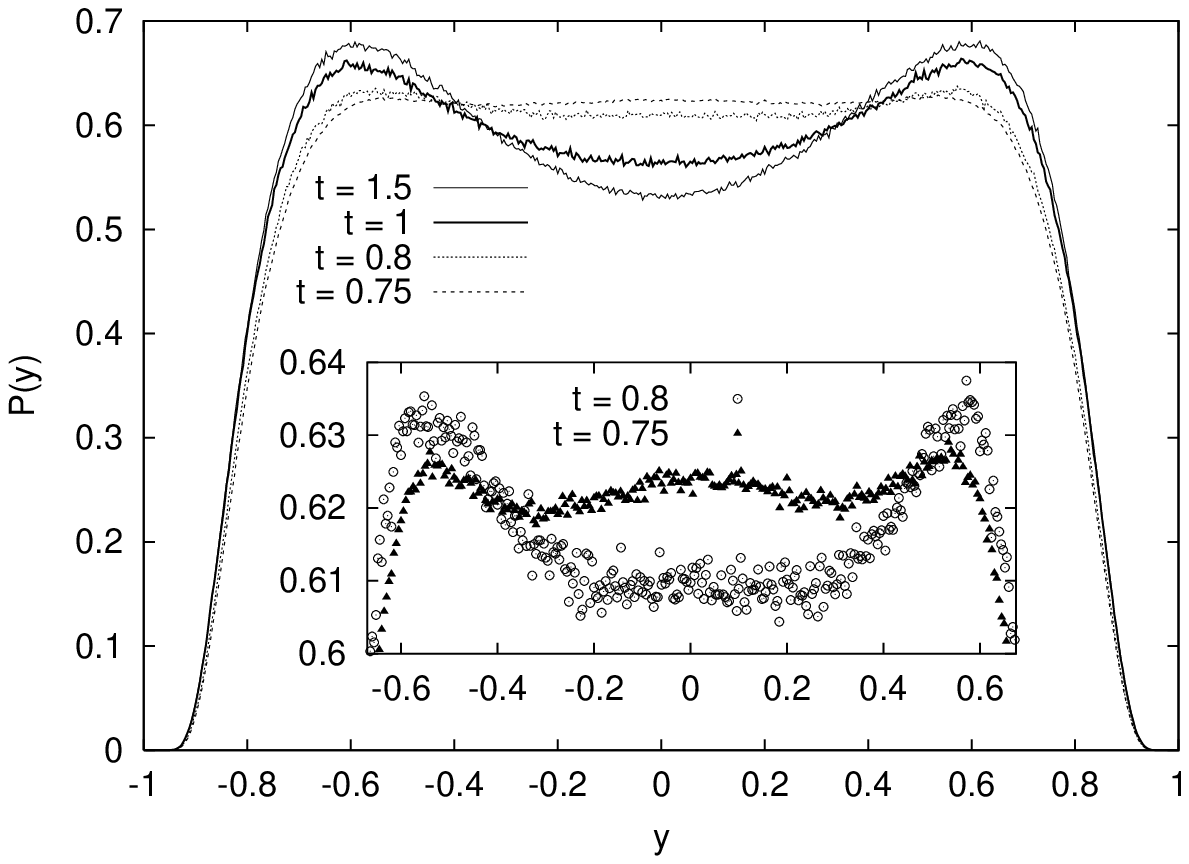,width=8.5cm,clip=}}
  \put(-210,-15){(b)} \caption{\label{fig:peaks} Distribution function
    for the projection of the free end along the transverse
    direction, $P(y)$, obtained by Monte Carlo simulations.  Lengths
    are measured in units of $L$. Errors are comparable to the point
    size in the insets. (a)~Appearance of double peaks for $t \lesssim
    2.5$. (b)~Re-entrance from the double peaks to a flat distribution 
    in the stiff limit $t \lesssim 0.75$. Insets show details of the
    crossover regions.}
\end{figure}
As $t$ approaches the value $1$ from above (flexible side), the 
Gaussian peak is first smeared out into an intrinsically 
non-Gaussian flat distribution (see Fig.~\ref{fig:peaks}a). At 
$t=2.8$ (see inset), the distribution contains three local maxima, 
but as $t$ is decreased, the central peak at $y=0$ loses weight to 
the two symmetric peaks off the x axis. The double-peak structure 
is most pronounced around $t\approx 1.5$, i.e. $L \approx 1.5 
\ell_p$. 

As the stiffness is increased, $P(y)$ recovers its flat structure, 
as shown in Fig.~\ref{fig:peaks}b. Notice also (inset of 
Fig.~\ref{fig:peaks}b) that at $t=0.75$ the two peaks start to 
compete with a growing peak centered at $y=0$, such that one finds 
a triple maxima shape again. Although intrinsically non--Gaussian, 
this central peak will eventually tend to a Gaussian distribution 
in the stiff limit.  The re-entrance from the double peak 
structure to a flat distribution is a genuine hallmark of 
semiflexibility. This effect cannot be explained by analytical 
calculations using a harmonic (or weakly bending rod) 
approximation, whose prediction for $P(y)$ would be a Gaussian 
centered at $0$~\cite{wilhelm:unpub}. Higher order cumulant 
expansions about a Gaussian distribution have also failed to 
provide a fast convergence to our $P(y)$. An entirely analytical 
solution can be provided by the eigenfunction approach described 
in~\cite{bra03} for persistent random walks, although the 
connection to our probability distributions would only be 
numerical. 

Finally, let us emphasize that the double-peak structure of $P(y)$ 
does not indicate a bistability in the constant force ensemble. As 
shown below, linear response theory leads to positive force 
constants in this regime. What actually happens under the 
application of an external force is that the distribution function 
becomes asymmetric and weight is shifted from one peak to the 
other. In an experimental setting with a fixed transverse distance 
$y$ and a correspondingly adjusting force, one would probe $P(y)$ 
directly and be able to observe a kind of ``bistability''. 

\begin{figure}[htb!]
  \epsfig{figure=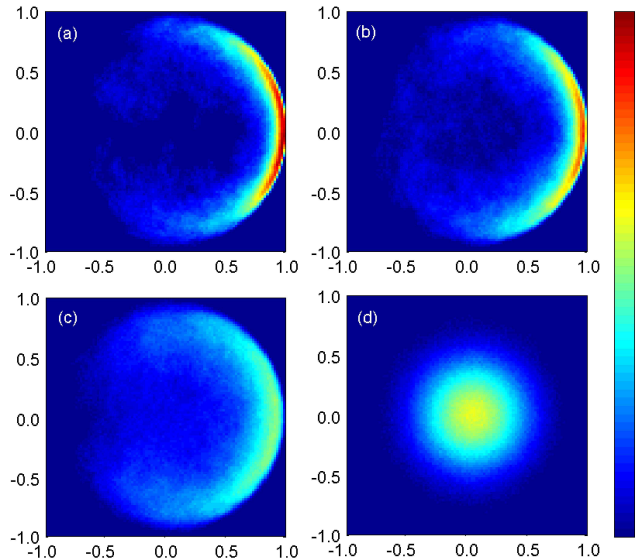,width=8.5cm,clip=}
\caption{\label{fig:dens} Density plots obtained by Monte Carlo 
  simulations: dense regions are colored in red, scarcely populated in
  blue on a color scale appropriately chosen to enhance the contrast.
  (a)~$t=2/3$; (b)~$t=1$; (c)~$t=2$; (d)~$t=20$.}
\end{figure}

Further insight can be gained by the inspection of the joint
distribution function $P(x,y)$, represented with density plots in
Fig.~\ref{fig:dens}.
In the stiff limit, $P(x,y)$ should be confined to the classical
contour obtained by applying the elasticity equations to a rigid 
rod. This contour can be approximated by a parabola in the 
proximity of full stretching and is obtained through elliptic 
functions for any deformation~\cite{lan7}. In Fig.~\ref{fig:dens}a 
the classical contour coincides with the ridge of the probability 
distribution function. As we relax the stiffness, thermal 
fluctuations will make the tip of the filament explore the 
configuration space in the vicinity of the classical contour.  
Roughly speaking, transverse (bending) fluctuations enhance 
fluctuations along the classical contour and shift weight from the 
center to the upper and lower wings in Fig.~\ref{fig:dens}a--b.  
In contrast, longitudinal fluctuations widen the distribution 
function perpendicular to the classical contour. Since for a 
semiflexible polymer, the corresponding lengths $L_\parallel$ and 
$L_\perp$ scale differently (transverse fluctuations are much 
``softer'' than longitudinal ones), upon lowering the stiffness 
$P(x,y)$ gains more weight in the wings rather than in the center.  
It is precisely this effect that gives rise to the double peak 
distribution, when $P(x,y)$ is projected in the transverse 
direction (see Fig.~\ref{fig:dens}b).  Eventually, in the flexible 
limit, where transverse and longitudinal fluctuations become 
comparable, $P(x,y)$ is spread so as to cover almost all the 
available space (Fig.~\ref{fig:dens}c), before the isotropic 
Gaussian distribution is recovered (Fig.~\ref{fig:dens}d). 

We have also explored the transverse response of semiflexible 
polymers by applying a constant force $f$ in the transverse 
direction. The effect of a small applied force on the average 
end-to-end distance (or force extension relation) has been studied 
within linear-response in~\cite{kro96}.  In this work, we will 
consider the effect of an external transverse force of arbitrary 
magnitude on the average position $\left< x \right>_f$ and $\left< 
y \right>_f$ of the free end. 

\begin{figure}[htb!]
  \epsfig{figure=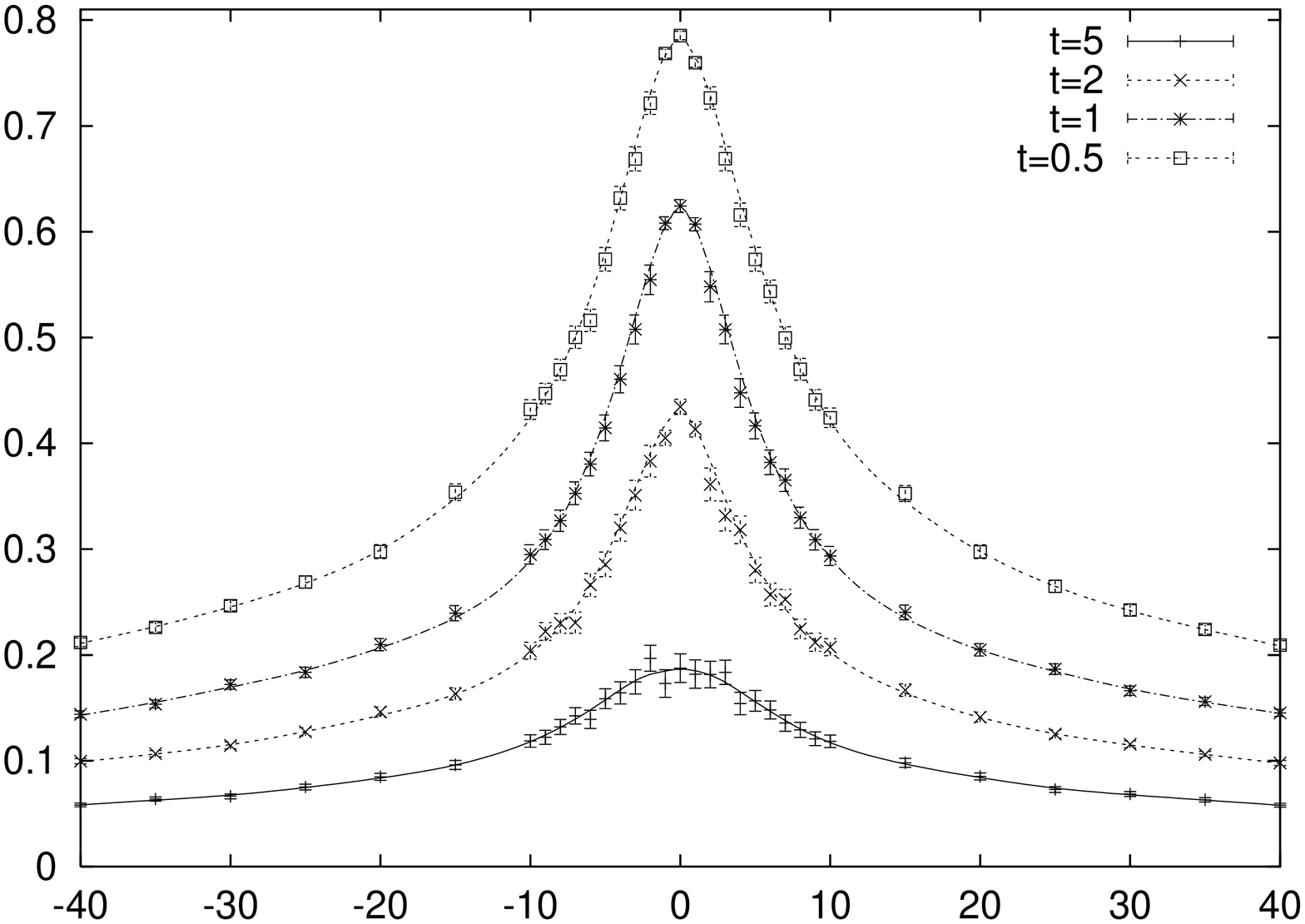,height=8.5 cm,angle=270,clip=}
  \put(-118,-175){$f$} \put(-253,-13){$\left< x \right>_f$}
  \put(-242,-175){\epsfig{figure=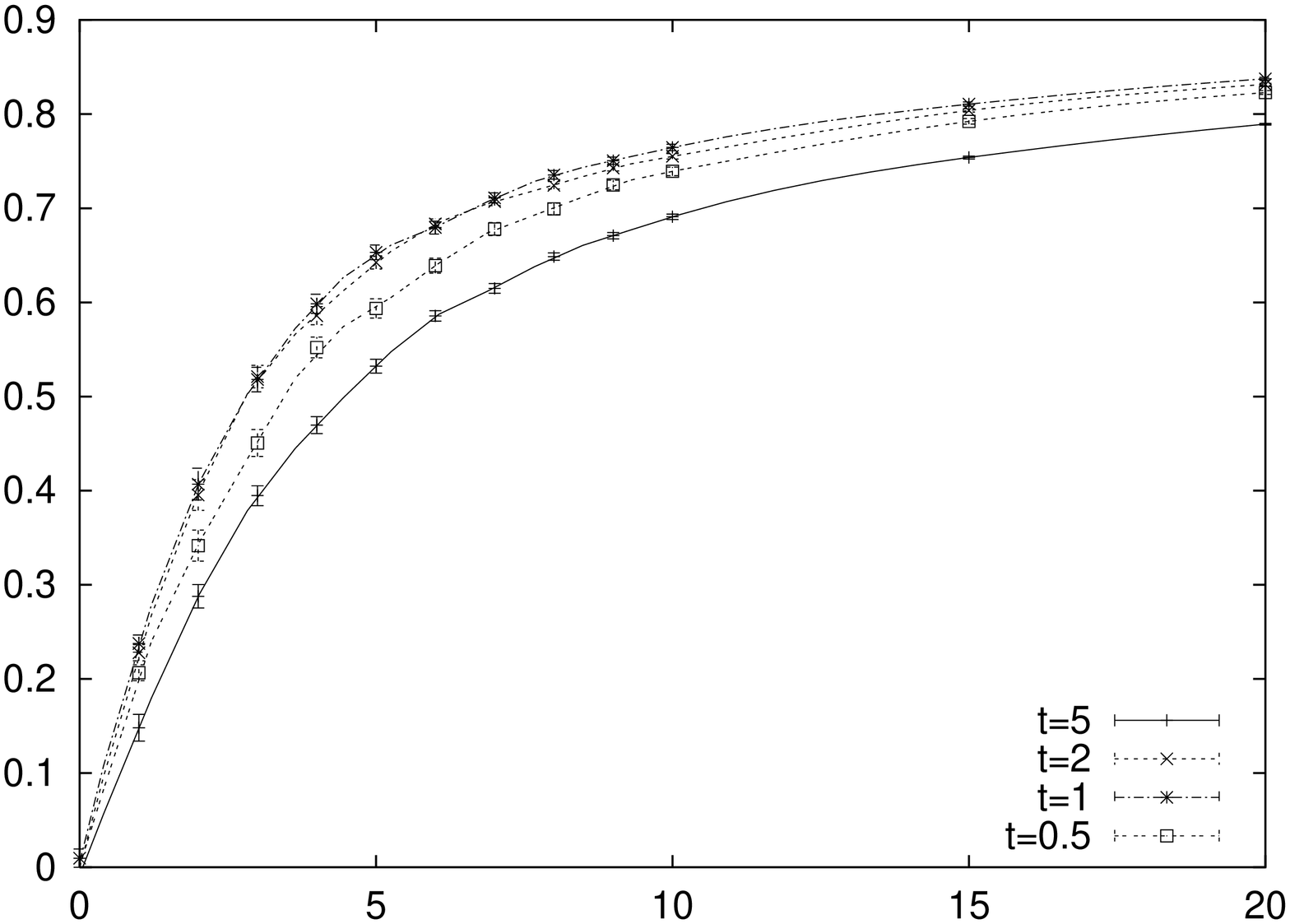,height=8.5
      cm,angle=270,clip=}} \put(-118,-350){$f$}
  \put(-253,-188){$\left< y \right>_f$} \caption{\label{fig:f_ext}
    Response to a transverse force, obtained by Monte Carlo
    simulations.  Forces are measured in units of $k_B T / L$, lengths
    in units of $L$. Error bars are shown. (Above) Response in the
    clamping direction. (Below) Response in the transverse direction
    is odd with $f$: only part of the explored parameter region is
    shown for clarity.}
\end{figure}

In general, we expect $\left< y \right>_f$ to have the same parity 
of the applied force, and hence to be odd, while $\left< x 
\right>_f$ should not depend on the sign of the force and hence 
should be even. In the continuum limit, it is possible to write 
down the exact expressions for $\left< x \right>_f$ and $\left< y 
\right>_f$ and to show that the expected parities hold on very 
general grounds and that  the response of the longitudinal 
extension to a transverse force is intrinsically nonlinear in the 
small force regime. Monte Carlo simulations confirm these 
predictions, as shown in Fig.~\ref{fig:f_ext}. The response in the 
direction of the clamped end is even in $f$ and it can be 
approximated by a parabola centered on the $f=0$ axis. The 
response in the transverse direction is odd in $f$ and shows the 
same re-entrance phenomenon reported in~\cite{kro96} for the 
linear response coefficient. 

Note that while in the case of a longitudinal force, the approach
towards full stretching (or saturation) can be calculated within 
the weakly bending rod approximation, this is no longer true for 
transverse forces. The position of the free end can be calculated 
from classical elasticity theory~\cite{lan7} and expressed by 
means of elliptic functions. Only in the high force regime or in 
the stiff limit, when fluctuations become unimportant, results 
from our simulations coincide with classical elasticity theory.

The effects hereby reported are amenable to a direct comparison 
with experiments regarding cytoskeletal filaments, or even DNA.  
For instance, optical systems might be used to get the $x$ or $y$ 
projection of the radial distribution function for a particular 
class of semiflexible polymers. For F-Actin with $\ell_p \approx 
16 \ \mu$m~\cite{osk02}, the double peak effect should be well 
visible for a range of lengths, $ 12 \ \mu\mbox{m} \lesssim L 
\lesssim 43 \ \mu\mbox{m}$. In this parameter range the difference 
between the central relative minimum and the double peaks maxima 
results in $10\%$ of the total length (see Fig.~\ref{fig:peaks}), 
in the range $1\div4 \ \mu$m that is well above the experimental 
precision of $0.05 \ \mu$m reported by~\cite{osk02}.  Hence 
F-Actin would provide an ideal benchmark for the effects we 
report. We emphasize that the double peak structure is a clear 
hallmark of semiflexibility and hence it might be used to obtain a 
rough estimate of the persistence length of a particular polymer 
filament, as for instance the nanometer sized stalks of kinesins 
and myosins. 

In summary, we have presented evidence from extensive Monte Carlo 
simulations that the parameter region corresponding to 
semiflexible polymers is hallmarked by the appearance of a series 
of effects in the radial distribution function and in the response 
of the clamped polymer to an external transverse force. A 
semiflexible polymer shows a distinct anisotropy in the 
probability distribution function of the free end along the 
direction of the clamped end. At intermediate stiffness, $L\approx 
\ell_p$, the distribution function shows a pronounced double peak 
structure in the transverse direction. Semiflexible polymers have 
been previously reported~\cite{kro96} to be anisotropic objects, 
i.e. to respond in different ways to forces applied in the 
clamping or transverse direction. Here we have shown that even 
their response to a force along the transverse direction alone is 
intrinsically anisotropic, being linear in the transverse 
direction and non-linear along the direction of the clamped end in 
the small force regime. 

We acknowledge helpful discussions with P.~Benetatos, 
A.~Parmeggiani, J.~Wilhelm, T.~Franosch and K.~Kroy. This research 
has been supported by a Marie Curie Fellowship under contract no. 
HPMF-CT-2001-01432.

\end{document}